\documentclass[%
 aip,
 amsmath,amssymb,
 reprint,%
]{revtex4-1}

\usepackage{graphicx}% Include figure files
\usepackage{dcolumn}% Align table columns on decimal point
\usepackage{bm}% bold math
\usepackage[utf8]{inputenc}
\usepackage[T1]{fontenc}
\usepackage{mathptmx}
\usepackage{etoolbox}
\usepackage{amssymb,xcolor}

\makeatletter
\def\@email#1#2{%
 \endgroup
 \patchcmd{\titleblock@produce}
  {\frontmatter@RRAPformat}
  {\frontmatter@RRAPformat{\produce@RRAP{*#1\href{mailto:#2}{#2}}}\frontmatter@RRAPformat}
  {}{}
}%
\makeatother
\begin{document}

\preprint{AIP/123-QED}

\title[]{Thresholded quantum LIDAR in turbolent media}
% Force line breaks with \\

\author{Walter Zedda}
%\email{wal.zedda@stud.uniroma3.it}
\affiliation{Dipartimento di Matematica e Fisica, Universit\`a degli Studi Roma Tre, Via della Vasca Navale 84, 00146 Rome, Italy}

\author{Ilaria Gianani}
\email{ilaria.gianani@uniroma3.it}
\affiliation{Dipartimento di Scienze, Universit\`a degli Studi Roma Tre, Via della Vasca Navale 84, 00146 Rome, Italy}

\author{Vincenzo Berardi}
\affiliation{Dipartimento Interateneo di Fisica ``Michelangelo Merlin'', Politecnico di Bari, Via Orabona 4, 70126 Bari, Italy}

\author{Marco Barbieri}%
%\email{marco.barbieri@uniroma3.it}
\affiliation{Dipartimento di Scienze, Universit\`a degli Studi Roma Tre, Via della Vasca Navale 84, 00146 Rome, Italy}
\affiliation{Istituto Nazionale di Ottica, CNR, Largo Enrico Fermi 6, 50125 Florence, Italy}

\begin{abstract}
Light detection and ranging is a key technology for a number of applications, from  relatively simple distance ranging to environmental monitoring. When dealing with low photon numbers an important issue is the improvement of the signal-to-noise-ratio, which is severely affected by external sources whose emission is captured by the detection apparatus. In this paper, we present an extension of the technique developed in [Phys. Rev. Lett. 123, 203601] to the effects caused by the propagation of light through a turbulent media, as well as the detection through photon counting devices bearing imperfections in terms of efficiency and number resolution. Our results indicate that even less performing technology can result in a useful detection scheme.

\end{abstract}

\maketitle

%\pacs{}% insert suggested PACS numbers in braces on next line

\maketitle

%\tableofcontents

\section{\label{Introduction}Introduction}

LIght Detection And Ranging (LIDAR) is one of the most exploited techniques for remote investigations, such as atmospheric monitoring \cite{Weibring:03}, laser ranging \cite{Riemensberger2020}, and pollution control \cite{NAP18733}. As recurrent for many sensing and communication technologies, the quest for improved performances has fueled the effort toward schemes encompassing quantum effects. In particular, this has stimulated a quest for implementations of LIDAR and cognate measurement techniques down at the single-photon level~\cite{Gariepy:2015cr,Tachella:2019dq,Li:20,Li:21,Rapp:21,radarreview}.

By its nature, LIDAR is operated in the presence of conspicuous loss, as the signal can propagate for great distances and in noisy environments.  In these conditions, quantum illumination has been identified as an intriguing option~\cite{science.1160627,PhysRevLett.101.253601,PhysRevLett.110.153603}, but its effectiveness is currently a matter of debate~\cite{Nair:20,9087936}. In fact, it occurs that the control of the quantum state of light may provide little advantage~\cite{PhysRevA.80.063803,PhysRevLett.102.040403}. The work of Cohen and coworkers~\cite{Cohen-2019} has demonstrated that improvement can be attained by accessing quantum properties at the detection level. In particular, they have revealed that superior performance can be obtained in a time-of-flight ranging measurement at low signal and high noise by applying a threshold $S$ to photon number detection. The enabling mechanism can be traced in the different photon statistics of the signal - a coherent state - and the noise - a thermal background.

Detectors that genuinely resolve the number of arriving photons have represented a major breakthrough for quantum technologies~\cite{irwin95,Cabrera:1998,PhysRevA.71.061803,Smith:2012yu,Hum:2015,Hopker:19}. Their operation, at present, is limited by the required cryogenic temperature; their inclusion in LIDAR schemes seems to privilege solutions pertinent to stable measuring stations, but they have reduced appeal for portable devices. In this respect, a different solution, based on multiplexing of avalanche photodiodes~\cite{PhysRevA.67.061801,PhysRevA.68.043814,jmodo04,afek09}, could be more interesting from a technological perspective, though it should be borne in mind that the observed count statistics only resembles the actual photon statistics, but it is not strictly equivalent to it~\cite{PhysRevLett.109.093601,PhysRevA.95.023815,PhysRevA.99.043822,bjork20}. The relatively slow time resolution of these detectors makes them suitable for ranging over great distances when employed for time-of-flight measurements. Over these lengths, turbulence, giving origin to scintillation~\cite{1451964,Milonni_2004,Dios:04,PhysRevA.80.021802,PhysRevLett.109.200502,Bohmann:17}, modifies the photon statistics of the signal. All these effects impact the elementary functioning of the detection scheme, and thus need being scrutinised.

In our work we discuss the model of a portable LIDAR device operating in the long-distance regime. This builds on the original scheme of~\cite{Cohen-2019}, but includes two important variations. We consider the unavailability of real photon-number detectors, hence replacing them with multiplexed detectors. Further, we include the effect of scintillation on the photon statistics, affecting the basis of the discrimination mechanism.

The performance of this scheme is robust against the limiting factors mentioned above. For its assessment, we do not rely uniquely on the signal-to-noise ratio, but consider instead the number of copies needed for a reliable identification of the signal.

%, and a careful choice of the threshold parameter $S$  allows to disentangle the scintillation contribution from the noise. This is the critical parameter needed to limit the impact of turbulence. 

\section{\label{Results}Results}

\begin{figure}[h]
\includegraphics[width=\columnwidth]{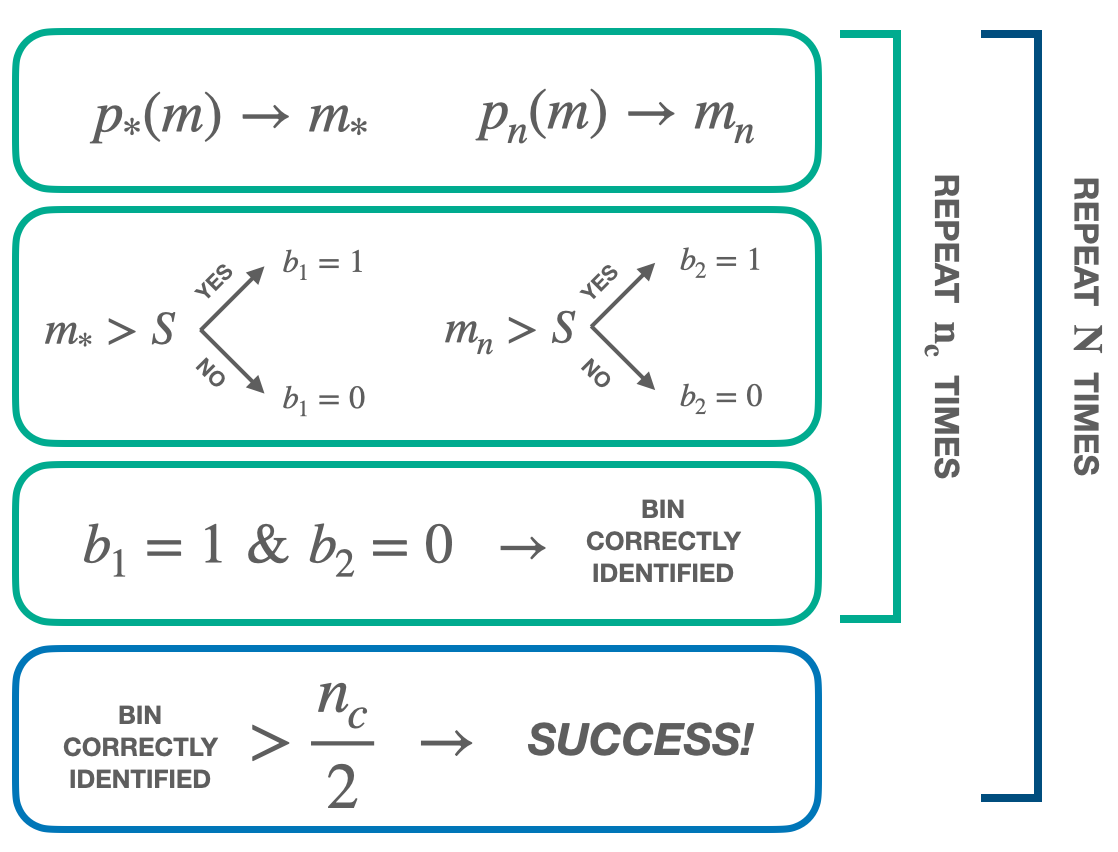}
\caption{Scheme of the simulated process adopted for the evaluation of the probability of successful runs.}
\label{fig:ilsetti}
\end{figure}

The sought application is quantifying the distance of a remote object by means of time-of-flight measurement of reflected light. A pulsed laser beam, characterised by an average number of photons per pulse, $\mu_{s}$, travels towards a reflecting surface (target), where it is reflected back to a photon counting detector. We consider ideal scenario where all the photon reaching the target are reflected by its surface. The detection is affected by noise due to background light with mean photon number $\mu_{n}$. Depending on the optical bandwidth of the detection, this can either be described by a single thermal mode, or by a collection of many modes that give rise to Poisson photon statistics. The problem is thus the old one of separating the signal from the noise in the photon counting realm.

We can consider an ideal detector as a device able to discriminate the photon number in each pulse with unit efficiency, $\eta=1$, hence no dark counts, {\it i.e.} the associated probability $p_d$ is set to zero. While this is an extreme idealisation, a similar behaviour is observed in superconducting transition edge detectors~\cite{Lita:08,Humphreys_2015,Hopker:20}. These are bolometers that can operate with efficiency exceeding $95\%$. A less demanding solution is represented by the adoption of multiplexed avalanche photodiodes (M-APDs).  An APD is sensitive to a single photon, however its response, a `click', is independent on the actual number of photons hitting its active area. If light is divided into $M$ modes before detection and each is monitored by an APD, then the distribution of clicks from the whole detection system gives partial information about the photon distribution. In the limit of infinite multiplexing, the exact photon distribution can be recovered, though with a slow convergence going as $1/M$~\cite{PhysRevA.85.023820,PhysRevLett.109.093601}. The difference between click and photon distribution thus remains sizeable for $M\sim10$, which are typically adopted~\cite{PhysRevLett.110.173602}. The relevance of multiplexed detection for LIDAR must then be assessed based on its actual response.

For our application, the detector operates in a triggered mode from a fast photodiode signal monitoring the laser pulses. For a single device, the time resolution is dictated by its jitter, since this determines the time bin width. In a multiplexed device, two aspects are to be considered: the synchronisation of the individual detectors and the individual jitter. The former can be tackled by means of the reading electronics and typically poses little challenges. In these conditions, the width of the time bin is set mainly by the element with the largest jitter. Thus, the time resolution is brought from $T_d=30$ns (the typical dead time for an APD in free running) down to $T_j=300$ps (the time jitter of the APD). This  limits the uncertainty on the distance within $cT_j\simeq 10$cm at best. 

%For our application, we imagine that the time resolution is dictated by a binning of the arrival times, as densely as the response of the detector allows. 

Given that pulses can be made shorter than this limit, the problem can be formulated as the one of identifying the one bin in which the signal accumulates against the others, only populated by the noise. The noise is  approximately the same in all bins - including the one with the signal.The good bin will then accumulate photons coming from both signal and noise, whereas all others (bad bins) will only accumulate background noise. 

We consider here the illustrative case of a binary decision problem and consider only two bins. Our aim is that of establishing if we can correctly identify the bin where both the signal and noise are detected against the bin receiving only noise. The output of our detection is reduced to a yes-no answer based on a threshold $S$ to the number of detected events. We first deal with the ideal case of a photon-number resolving detector. In the bad bin, the probability for an $m$-photon detection event is given by $p_{n}(m)$. This can be taken as a thermal distribution $p_{ n}(m)=\mu_n^m/(1+\mu_n)^{1+m}$ if a single mode is detected. In the opposite limit, if a large number of modes are collected, the resulting multi-thermal statistics becomes Poissonian $p_{ n}(m)=e^{-\mu_n}\mu_n^m/m!$~\cite{PhysRevLett.101.053601}. 

In the good bin, on the other hand, where signal and noise coexist, an $m$-photon detection event is given by the convolution of the distribution of the signal $p_{s}(m)$ with the distribution of the noise $p_{n}(m)$: 
\begin{equation}
\label{eq:mixture}
    p_{*}(m)=\sum_{l=0}^{m} p_{s}(m-l)p_{n}(l),
\end{equation}
because the two distributions are independent. The signal distribution $p_s(m)$ from a laser is Poissonian, but this is modified in the presence of atmospheric disturbance, such as scintillation.  

%whose  width is determined by the response time of the detector.
%A suitable figure of merit for assessing the performance of our scheme is thus to evaluate 

For our analysis we choose as a figure of merit the probability $P$ of correctly identifying the good bin as a function of the number $n_c$ of repetitions. This is a different approach than the one in~\cite{Cohen-2019}, which instead focused on the signal-to-noise ratio of the measurement. For perfect number-resolving detectors, our simulations proceed as illustrated in Fig.~\ref{fig:ilsetti}
\begin{enumerate}
    \item we generate an event with $m_*$ photons extracted from the probability distribution $p_*(m)$ in \eqref{eq:mixture} associated to the first time bin;
    \item similarly we generate an event with $m_n$ photons extracted from the probability distribution $p_n(m)$ associated to the second time bin;
    \item we set the value $S$ as the threshold, which will govern the performance of the protocol;
    \item we then assess whether $m_*>S$, if so, we tag the first bin as $b_1=1$, otherwise $b_1=0$;  
    \item we carry out the same comparison for $m_n > S$ to define the tag for the second bin $b_2$;
    \item the steps 1-5 are iterated $n_c$ times. If the majority of the iterations return $b_1 =1$ and $b_2=0$, then the run is considered successful; otherwise either we incorrectly identified bin n.2 as the good one, or obtain an inconclusive result;
    \end{enumerate}
Finally, the probability of success can be evaluated by repeating the procedure $N$ times and calculating the frequency of the successful runs $N_s$
\begin{equation}
    P=\frac{N_{s}}{N}.
\label{eq:prob}
\end{equation}

The same analysis can be carried out for realistic detectors with an important difference: the events of time bin n.1 and n.2 are drawn by the click statistics from the M-APDs. Their distribution can be derived from $p_*(m)$ and $p_n(m)$ following the methods illustrated in~\cite{bjork20}, that also discusses how to include dark counts occurring with probability $p_d$. Apart for this change, the procedure is unaltered.

%in this case, the outputs are the number of clicks originating from the M-APDs, to which the threshold $S$ is applied. 

    %for each detection event, count the number, $n_{m}$, of photons coming from the signal+noise distribution above $S$, as well as the number, $n_{t}$, of photons coming from the noise distribution above $S$;

    %\item if only bin n.1 has $n_m\geq S$ the result of the detection event is a success, that is we can safely identify  bin n.1 as the one populated by the signal; otherwise,  we failed at identifying the good bin correctly, either because we have mistaken the bad bin for the good one, or because we have an inconclusive event.

We will show here four different cases:
\begin{itemize}
    \item High Signal - Low Noise (HSLN): $\mu_{s}=10$, $\mu_{n}=1$;
    \item Low Signal - Low Noise (LSLN): $\mu_{s}=1$, $\mu_{n}=1$;
    \item High Signal - High Noise (HSHN): $\mu_{s}=10$, $\mu_{n}=10$;
    \item Low Signal - High Noise (LSHN): $\mu_{s}=1$, $\mu_{n}=10$.
\end{itemize}
These cover illustrative operation regimes. For each one of them, we then evaluate the probability $P$ of a successful identification on the good bin as a function of $n_{c}$ for both ideal and realistic detection schemes.

%If our hypotheses on the effect of atmospheric scintillation are valid, we do expect that there is a lower limit for $\sigma$, namely $\sigma_{0}$, indicating that the scintillation phenomenon can be safely neglected and the result we obtain by including the scintillation has to be consistent with those obtained in absence of scintillation. 

% Our simulations have shown that the upper limit is $\sigma_{0}=0.63$, corresponding to $SI_{0}=4.9x10^{-1}$, that is, if $\sigma_{0}\leq0.63$ we can assume that there is no atmospheric scintillation.

% We plot, given a fixed time interval, the normalised probability, $p$, of detecting $\mu_{s}$ photons as a function of the number of replicas, $n_{c}$, i.e. the number of laser pulses.

%\subsection{\label{ideal detector} Ideal detector}

We first consider the ideal detector (quantum efficiency $\eta=1$,  no dark counts noise, $p_d=0$), which is able to count each and every photon impinging on it. This corresponds to the detection scheme introduced in~\cite{Cohen-2019}, analysed according to our figure of merit. 

We center our discussion on how different choices of $S$ influence the performance of the detection system, in particular the number of repetitions $n_c$ which are needed for a near-certain identification of the good bin. Indeed, for a given repetition rate of the laser, a lower $n_c$ corresponds to a faster measurement; this is relevant when tracking moving objects.

\begin{figure*}[hbt]
\includegraphics[width=\textwidth]{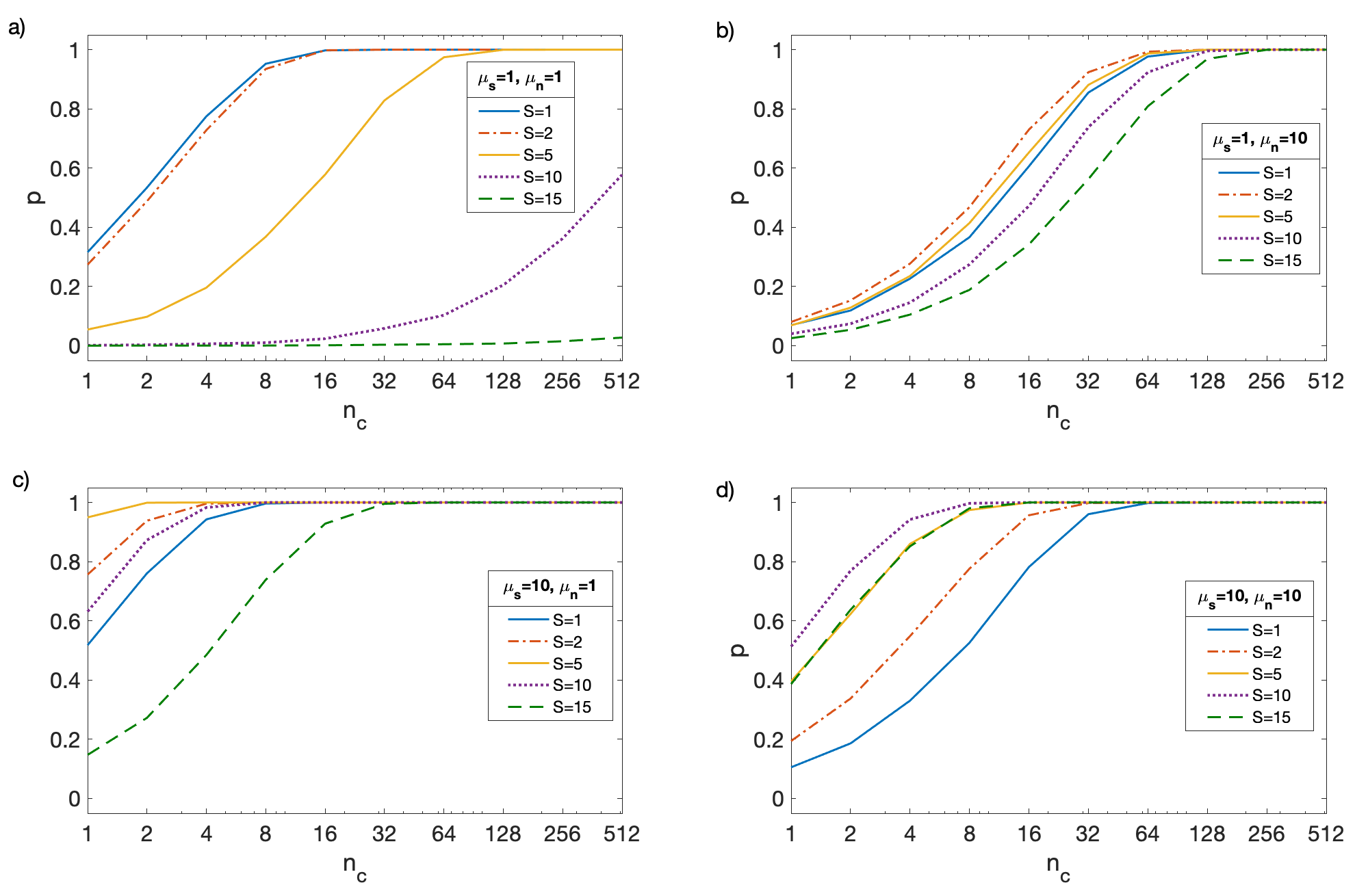}
\caption{Probability $P$ of correctly identifying the time bin populated by the signal, as a function of the number $n_c$ of pulses for an ideal detector. The different curves correspond to distinct thresholds. 
The four different panels refer to the four signal-noise conditions introduced the main text: a) LSLN, b) LSHN, c) HSLN, and d) HSHN.}
\label{fig:01}
\end{figure*}

Fig.\ref{fig:01} shows the probability $P$ vs. $n_{c}$ in Eq.\eqref{eq:prob} from our simulations for the case of a Poissonian signal and thermal noise.  Our goal is to set the threshold $S$ to a satisfactory value for which $P\simeq1$ is reached for the minimum value of $n_{c}$. As expected the value of $S$ depends on the choice of $\mu_{s}$ and $\mu_{n}$. Specifically, we find:
\begin{itemize}
    \item LSLN - $S=1$, $n_{c}=16$;
    \item LSHN - $1<S<5$, $n_{c}=64$;
    \item HSHN - $S=10$, $n_{c}=8$;
    \item HSLN - $S=5$, $n_{c}=2$.
\end{itemize} 
Besides for the case of LSHN where the process is obviously noise driven and we need a remarkably high number of detection events ($n_{c}$) to identify the signal, it is worth nothing that in all other three cases we can get $P\simeq1$ for $n_{c}=16$ at most. Further simulation considering multithermal noise with the same average intensity does not notably affect our findings. 

In general, obtaining a value for $S$ analytically is complex even in this idealised case, as it would involve calculating cumulants of the probabilities $p_*(m)$. No analytical expression is known, requiring numerical methods. 

%\subsection{\label{real detector} Real detector}

\begin{figure*}[hbt]
\includegraphics[width=\textwidth]{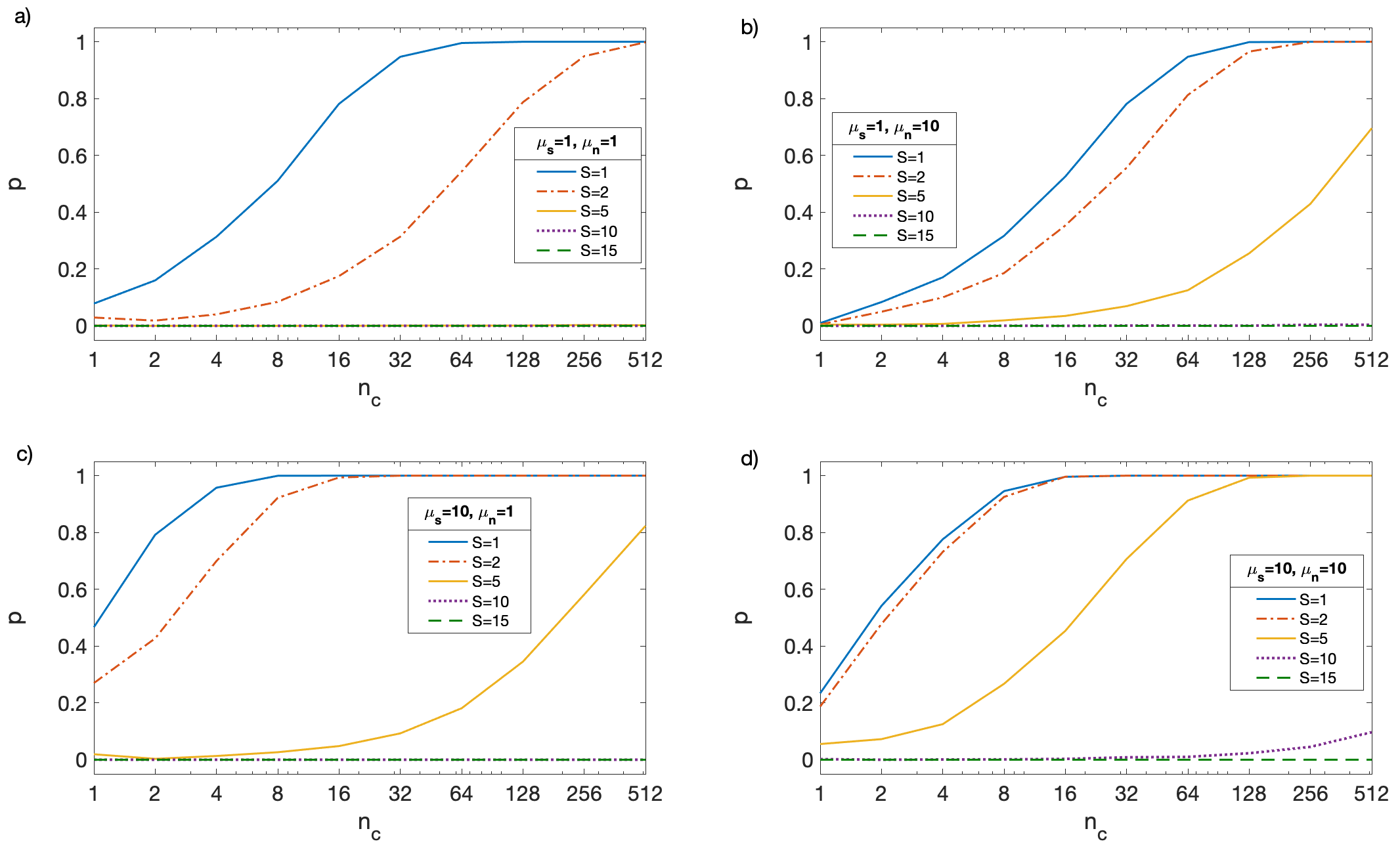}
\caption{Probability $P$ of correctly identifying the time bin populated by the signal, as a function of the number $n_c$ of pulses for a real detector. The different curves correspond to distinct thresholds. 
The four different panels refer to the four signal-noise conditions introduced the main text: a) LSLN, b) LSHN, c) HSLN, and d) HSHN. In all panels $\eta=0.1$, $p_d=1.0\,10^{-4}$}
\label{fig:03}
\end{figure*}

Having established the limits for ideal detectors, we now descend into the realm of real M-APDs. That is, we  consider that the quantum efficiency is lower, much lower, than $1$, that there is a certain number of dark counts, and that the photon number resolution is not perfect, but rather limited by multiplexing. In our simulations, we consider $\eta=0.1$ and $p_d=10^{-4}$, which are typical in the experiment, and we  multiplex over 16 equally probable APDs. The signal is in a coherent state, and the noise is thermal. Fig.\ref{fig:03} shows the results we obtained for different values of the threshold. Differently from the case of ideal detector, our results show that $S=1$ is always the best choice and that we get $P$ almost equal to $1$ for both HSLN and HSHN, whereas we need $n_{c}\geq 64$ for the  two cases in which the signal is low (LSLN, LSHN). This amounts to say that a simple click/no click scheme is optimal for the purpose of ranging under these conditions. Despite the many non-idealities of this scheme, it remains efficient since we have a mere four-fold increase in the number of required events for confident identification of the good bin.

%\subsection{\label{Appendix_scintillation}Atmospheric Scintillation}

During the back and forth trip, light passes through a turbulent medium, imposing scattering and phase shifts. This demands to take the phenomenon of atmospheric scintillation into account. Its usual treatment assumes that the transmission channel can be portioned into a large number of small homogeneous volume elements, which equally contribute to scattering and phase shifting~\cite{Zhu-2002}. The photon distribution of the laser will then be modified with respect to the initial Poissonian; since our measurements are phase-independent, we can focus on the implications for the intensity only.

Under the hypotheses above, one can invoke the central limit theorem to establish that the final distribution of the logarithm of the signal intensity $q$ can be assumed to follow a normal Gaussian distribution, whose width depends on the travelled distance~\cite{Zhu-2002}. The resulting distribution for $q$ is the log-normal~\cite{Milonni_2004}:
\begin{equation}
\mathcal{P}(q)= \frac{1}{\sigma q \sqrt{2 \pi}} e^{-\frac{(\log (q/\bar{q})+\frac{1}{2}\sigma^2)^2}{2\sigma^2}}
\label{eqn:milonni5}
\end{equation}
where $\bar{q}$ is the average number of counts, and $\sigma^2$ is the variance of $\log q$.

In the absence of background noise, Mandel's formula determines the probability of an $m$-photon event as
\begin{equation}
    p_s(m)=\left\langle \frac{q^m}{m!}e^{-q}\right\rangle
\end{equation}
where $q$ is the mean value of photon counts in the bin, and the average is taken over all possible turbulence configurations. Mandel's formula then writes
\begin{equation}
    p_s(m)=\int_0^\infty e^{-q}\frac{q^m}{m!}\mathcal{P}(q)dq.
\end{equation}
Finally, these probabilities are then convoluted with the noise by means of \eqref{eq:mixture} to obtain the actual distribution in the bin.
%It is customary to introduce the scintillation index $SI$, defined as the normalised variance of $q$:
%\begin{equation}
%SI= \frac{\bigl< (q-\bar{q})^2 \bigr>}{\bar{q}^2}
%\label{eqn:SI}
%\end{equation}
%This is connected to the width $\sigma$ of the photon distribution as
%\begin{equation}
%\sigma^2=\log (1+SI)
%\label{eqn:sigma}
%\end{equation}.

We can now proceed one step beyond and consider the effect of the different photon statistics from scintillation on our measurement. The time resolution is such that the signal is not spread across multiple bins, thus the binary decision approach we set initially remains valid. We first consider the ideal scheme, under the most favorable situation: low noise (LN) and either high signal (HS) or low signal (LS). The threshold $S$ is kept constant with respect to our previous analysis in order to isolate the implications of the variation of the distribution only.

\begin{figure*}[t]
\includegraphics[width=1\textwidth]{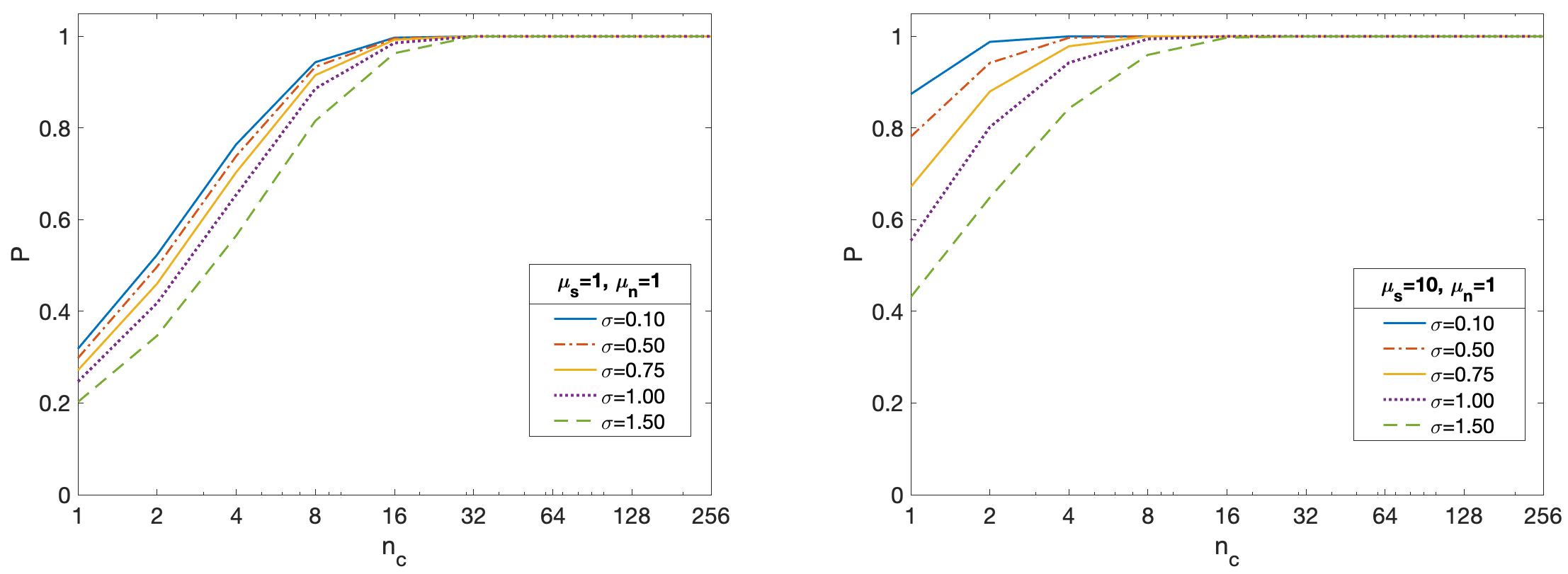}
\caption{Effect of scintillation on the success probability $P$ {\it vs.} $n_{c}$ for different values of $\sigma$ for ideal detectors for a) LSLN, b) HSLN.}
\label{fig:02}
\end{figure*}

As evident from Fig.~\ref{fig:02}, though somewhat expected, the modified statistics has limited implications in the case of LSLN, where $n_{c}$ tends to $128$ for all values of $\sigma$. When dealing with HSLN, this aspect of scintillation is more evident and in case of $\sigma=1.5$, we need $n_{c}$ as high as $16$ to obtain a good probability $P>0.9$ of identifying the good bin.

\begin{figure*}[ht]
\includegraphics[width=1\textwidth]{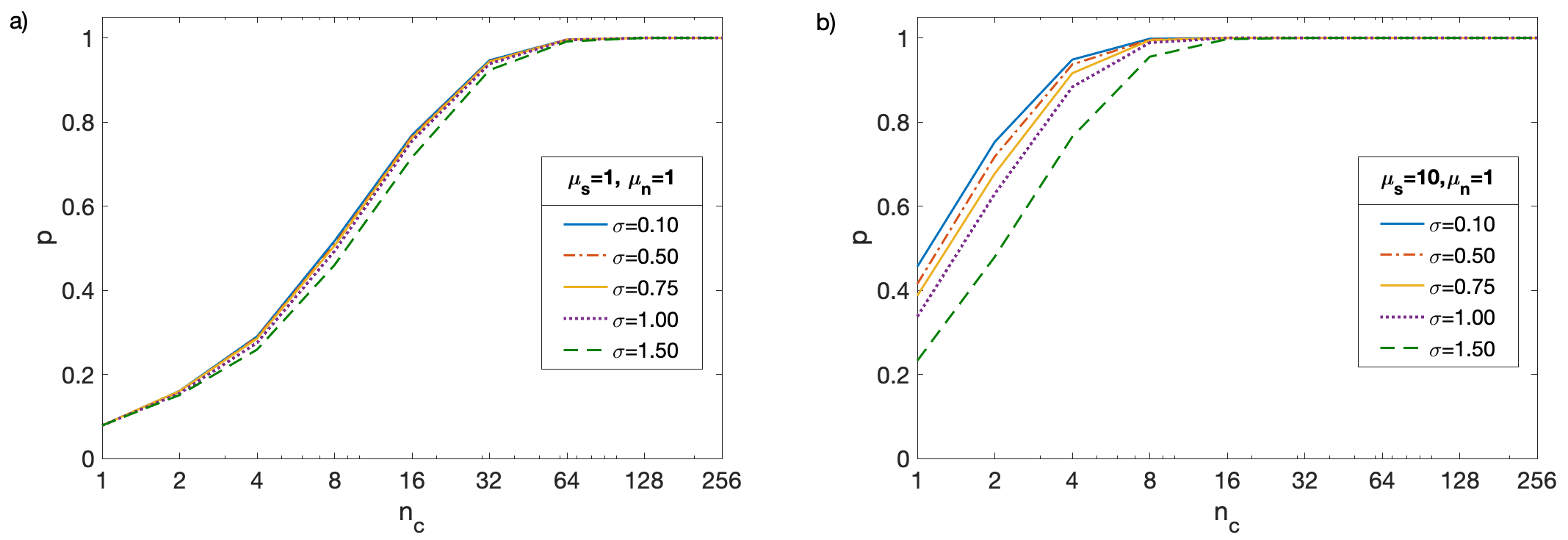}
\caption{Effect of scintillation on the success probability $P$ {\it vs.} $n_{c}$ for different values of $\sigma$ for real detectors for  a) LSLN; b) HSLN. In both panels panels $\eta=0.1$, $p_d=1.0\,10^{-4}$}
\label{fig:04}
\end{figure*}

Concerning the real detector, we restrict our analysis to the cases of HSLN and LSLN with $S=1$. Again, from Fig.~\ref{fig:04} it is clear that the change in statistics shows reduced effects, and confident discrimination can be achieved by a four-fold increase of the repetitions for the highest $\sigma$ we consider. 

It should be remarked that in all panels of Fig.~\ref{fig:02} and \ref{fig:04} the average flux $\bar q$ is held constant; this means that the higher loss from the turbulence are compensated by a higher intensity from the laser source. This increased demand in intensity is the main implication of scintillation.

%A more interesting situation occurs when the signal is considerably higher that the noise (HSLN). In this case by a careful choice of the threshold we can either operate approximately in a scintillation-independent mode by setting the threshold to $S=1$; this is equivalent to include the effect of scintillation into the noise. However, by raising $S$, for example setting $S=5$, we can actually disentangle the scintillation contribution from the actual noise by monitoring the value of $n_{c}$ needed to get to $P=1$. The lower the value of $n_{c}$ the more relevant the contribution of scintillation.

\section{\label{Conclusions}Conclusions}

In our analysis we examined the reliability of photon counting when dealing with LIDAR measurements in presence of atmospheric turbulence. We have first shown that in presence of an ideal photon counting device we need a fairly small number of detection events ($n_{c} \approx 8$) for an efficient signal detection. The introduction of a real detector increases such number to $16$ in case of high signal and low noise (HSLN) and to $32$ for the case of Signal and Noise comparably low (LSLN). For both ideal and real detectors, these values are not affected significantly by the modification of the photon statistics as the signal propagates through a turbulent medium. 
%However, when we study the behaviour of a real detector, even though we are able to work in a condition in which the scintillation can still be easily handled by a careful choice of the threshold $S$ and by a comparison of the results obtained for different values of $S$ (i.e. $S=1$ and $S=5$), we can study the effect of scintillation itself.

The same approach based on an hypothesis test lends itself to be applied to the spectral domain. This offers the chance to reveal the presence of gas species by assessing whether their specific Raman lines bear signal or just noise.

\begin{acknowledgments}
This work was supported by the European Commission through the FET-OPEN-RIA project STORMYTUNE (grant agreement no 899587). \end{acknowledgments}

\section*{Data Availability}
The data that support the findings of this study are available from the corresponding author upon reasonable request.

\section*{Author Declarations}
{\it Conflict of interest.} The authors have no conflicts to disclose.

\bibliography{mainbib}% Produces the bibliography via BibTeX.

% figures all grouped here

\end{document}